\documentclass[reprint,amsmath,amssymb, aps, prb]{revtex4-1}
\usepackage{graphicx}
\usepackage{caption}
\usepackage{subcaption}
\usepackage{epsfig}
\usepackage{color}
\usepackage{amsmath}
\usepackage{amsfonts}
\usepackage{mathrsfs}
\usepackage{txfonts}
\usepackage{color}

\usepackage{hyperref}

\newcommand{\bbr}{{\mathbf{r}}}
\newcommand{\bbx}{{\mathbf{x}}}
\newcommand{\bbz}{{\mathbf{0}}}
\newcommand{\bbq}{{\mathbf{q}}}

\begin{document}
\title{Hall Viscosity Revealed  via Density Response} 
\author{ Biao Huang}
\affiliation{Department of Physics, The Ohio State University, Columbus, OH 43210, USA}
\date{\today}

\begin{abstract}
Quantum Hall systems are recently shown to possess a quantity sensitive to the spatial geometry and topology of the system, dubbed the Hall viscosity $\eta_H$. Despite the extensive theoretical discussions on its properties, the question of how to measure $\eta_H$ still poses a challenge. In this paper, 
we present a general relation  between Hall viscosity and susceptibility for systems with Galilean invariance. Thus, it allows for determination of $\eta_H$ through density response signatures. The relations are verified in the integer quantum Hall example, and is further illustrated in an effective hydrodynamic analysis. Since the derivation is based on Kubo formulae and assumes no more than conservation laws and translational symmetry, the results are applicable to a wide range of systems. 
 \end{abstract}
\maketitle

\section{Introduction}
In recent years, there has been heated discussions on Hall viscosity $\eta_H$ (or ``Lorentz shear modulus'')\cite{avron}, the third viscosity coefficient unique in 2-dimensional isotropic, parity odd systems.  
It can be regarded as the Berry curvature in the parameter space of metric tensor\cite{avron}, and is proportional to the Wen-Zee shift\cite{read}, a quantity characterizing the system's spatial topology\cite{wenzee}.  

Besides the topological significance, the influence of Hall viscosity on various systems shows up in multiple ways.
As discovered recently,  for integer quantum Hall systems\cite{dencurv1} and  Laughlin wave functions\cite{dencurv2}, the Hall viscosity is related to the density response to the variation of the sample's scalar curvature. It acts as an anomolous force on vortices and changes their streamlines in an eulerian vortex fluid\cite{vortex}.   More generally, the Hall viscosity appears in the low-momentum expansion of Hall conductivity $\sigma_H(q)$ for Galilean invariant systems subject to inhomogeneous electric field\cite{emresp1,emresp2}. Clearly, studying these physical consequences not only broadens our perspetive on Hall viscosity, but also facilitates designing experiments to measure it.

Density response measurement  is a powerful and widely applicable tool to probe various systems. In recent works, many authors have related Hall viscosity to density response functions, i.e. the static structure factor \cite{dencurv1,dtson11}, using certain trial wave functions.  It is intriguing to see whether there are {\em universal} relations between density response functions and Hall viscosity.
In this paper, we relate the Hall viscosity $\eta_H$  to the $q^4$ term in the susceptibility $\chi(\bbq,\omega)$ for a general Galilean invariant system, thus allowing for measurements of Hall viscosity through density response experiments. 
We first derive the relation  in the microscopic level using  Kubo formulae, and verify the results in the integer quantum Hall example. Then we present an independent derivation  by applying effective hydrodynamics. The latter method has been adopted  \cite{hydrorev}  to analyze collective modes of fractional quantum Hall system at filling $\nu=1/3$, which gives qualitative agreement with experiment.\cite{collectivefqhe}
The two derivations give the same result in the long wave-length expansion.
Since both of the derivations are independent of specific wave functions,  type of interactions, filling factors, and etc., the results (\ref{svw}), (\ref{sv}) are valid for a general class of Galilean invariant systems.

\section{Kubo Formulae Method}
We first sketch the idea of relating viscosity and conductivity to susceptibility before the strict derivation.
In a viscous fluid, the stress tensor would respond to the velocity gradient $v_{\alpha\beta} = \frac{1}{2}(\partial_\alpha v_\beta + \partial_\beta v_\alpha)$
 through 
\begin{equation}\label{stressdef}
  \delta T_{\mu\nu} = \eta_{\mu\nu\alpha\beta} v_{\alpha\beta},
 \end{equation} 
 where $\eta_{\mu\nu\rho\sigma}$ are the viscosity coefficients. In three dimensions, rotation symmetry reduces the number of  $\eta_{\mu\nu\rho\sigma}$ to 2, corresponding to the usual shear and bulk viscosity
\begin{equation}\label{shearbulkvisc}
\eta^S(\delta_{\mu\alpha}\delta_{\nu\beta} + \delta_{\mu\beta}\delta_{\nu\alpha} - \frac{2}{d}\delta_{\mu\nu}\delta_{\alpha\beta}), \qquad
\zeta\delta_{\mu\nu}\delta_{\alpha\beta},
\end{equation}
where $d$ is spatial dimension.
 But in 2D, the in-plane isotropy allows for an additional one called Hall viscosity 
 \begin{equation}\label{hallvisc}
 \eta_H(\delta_{\nu\alpha}\epsilon_{\mu\beta} - \delta_{\mu\beta}\epsilon_{\alpha\nu}).
 \end{equation} 
Meanwhile, the density and current of a system would respond to potential and electric field perturbations
\begin{eqnarray}
\delta n(\bbq,\omega) &=& \chi(\bbq,\omega) \varphi(\bbq,\omega),\\
\delta j_\alpha(\bbq,\omega) &=& \sigma_{\alpha\beta}(\bbq,\omega) E_\beta(\bbq,\omega),
\end{eqnarray}
where  $\chi(\bbq,\omega), \sigma_{\alpha\beta}(\bbq,\omega)$ are the (charge-)susceptibility and conductivity respectively.
The key idea is that since in linear regime, the different  transport processes probe the same equilibrium property of the system, the  transport coefficients are all related to each other. Technically, the essential ingredients in the Kubo formulae for the transport coefficients are the correlators $\chi\sim \langle[n, n]\rangle_0$, $\sigma_{\alpha\beta} \sim \langle[ j_\alpha, j_\beta]\rangle_0$, and $\eta_{\mu\nu\alpha\beta} \sim \langle[ T_{\mu\nu}, T_{\alpha\beta}]\rangle_0$. For Galilean invariant systems, the current is related to momentum by $\mathbf{j}=\frac{e}{m}\mathbf{p}$. Thus, the particle and momentum conservation for the system at equilibrium,
\begin{eqnarray}\label{ncons}
&&
\partial_t n(\bbx,t) + \partial_\alpha  p_\alpha(\bbx,t) =0,\\
\label{pcons}
&&\partial_t p_\alpha(\bbx,t) + \partial_\beta T_{\alpha\beta}(\bbx,t) =  f_\alpha(\bbx,t)
\end{eqnarray}
 naturally provides a bridge among the transport coefficients of different types. Roughly speaking, by writing the conservation laws in Fourier space, we see the viscosity is related to the $q^2$ and $q^4$ terms in conductivity and susceptibility respectively.   Analogous ideas have been  adopted recently to discuss the unitary Fermi gases \cite{taylorranderia} and the conductivity-viscosity relations \cite{emresp2}, while in this paper we focus on the density response signatures for 2D systems subject to uniform magnetic field.

The Hamiltonian describing density coupling to external potential is $H_1= \int d^2 x \,  \hat{n}(\bbx) \varphi(\bbx,t) = \int \frac{d^2 q}{(2\pi)^2} \hat{n}(\bbq)\varphi(-\bbq,t)$, where $\hat{n}(\bbx) = \sum_{j=1}^N \delta(\bbx-\hat{\bbx}^j)$ is the density operator, with the Fourier transform $\hat{n}(\bbq) = \sum_{j=1}^N e^{-i\bbq\cdot\hat{\bbx}^j}$.  For a translational invariant system, the susceptibility\cite{pn} is given by
\begin{equation}\label{susceptibility}
\chi(\bbq,\omega) = -\frac{i}{\hbar} \int_0^\infty dt\,  e^{i\omega^+ t} \int d^2 x e^{-i\bbq\cdot\bbx} \langle[\hat{n}(\bbx,t), \hat{n}(\bbz,0)]\rangle_0.
\end{equation}
Here $\langle \dots \rangle_0$ means ensemble average in the fully interacting equilibrium system, and  $\omega^+ = \omega + i\epsilon$ with $\epsilon\rightarrow0$ ensuring that the perturbation adiabatically sets in.  On the other hand, for current response to electric field, the Kubo formula gives the conductivity\cite{mahan}
\begin{eqnarray}\nonumber
\sigma_{\alpha\beta}(\bbq,\omega) &=& \frac{ie^2\bar{n}}{m\omega^+} \delta_{\alpha\beta}+ \frac{1}{\hbar\omega^+} \int_0^\infty dt\, e^{i\omega^+ t}   \\ \label{conductivity}
&& \qquad \times \int d^2 x \, e^{-i\bbq\cdot\bbx} \langle[ \hat{j}_\alpha(\bbx,t), \hat{j}_\beta(\bbz,0)]\rangle_0,\qquad
\end{eqnarray}
where $\bar{n} = N/V$ is the average particle number density.
Here the current density  $\hat{j}_\alpha (\bbx) = \frac{e}{2m} \sum_{i=1}^N \{ \hat{\pi}_\alpha^i, \delta(\bbx-\hat{\bbx}^i)\}$, where $\{\dots\}$ is anti-commutator, and $\hat{\pi}^i_\alpha = \hat{p}^i_\alpha - e\hat{A}_\alpha^i   (\hat{\bbx}^i)$.

Here we  follow the Kubo formulae construction in \cite{emresp2} and obtain the {\em frequency-dependent}  conductivity-viscosity relation in low momentum expansion up to $q^2$ term, valid for Galilean invariant systems,
\begin{eqnarray}\nonumber
&& \sigma^{(2)}_{\mu\nu}(\bbq,\omega) = \frac{e^2}{m^2(\omega^2-\omega_c^2)^2}  \\ \nonumber
&&\,\,\times \left. [(\zeta(\omega)- \frac{\kappa^{-1}_{int}}{i\omega})
\left(
\omega^2 q_\mu q_\nu + i\omega\omega_c q^2\epsilon_{\mu\nu} 
+ \omega_c^2 \epsilon_{\mu\beta}\epsilon_{\nu\delta} q_\beta q_\delta
\right) 
\right.
 \\ \nonumber
\\ \nonumber
&&  \quad + \eta^S(\omega) q^2 \left( (\omega^2+\omega_c^2)\delta_{\mu\nu} + 2i\omega\omega_c\epsilon_{\mu\nu}\right)\\  
\label{cv}
&& \quad \left. - \eta_H(\omega)\left( 
i\omega\omega_c(q_\mu q_\nu + q^2\delta_{\mu\nu}) 
- (\omega^2+\omega_c^2) q^2\epsilon_{\mu\nu}
\right)
\right. ],
\qquad\qquad
\end{eqnarray}
where $\omega_c=eB/m$ is the cyclotron frequency, 
and  the ``internal compressiblity'' is
\begin{equation}
\kappa_{int}^{-1} = -V\left(
\frac{\partial P_{int}}{\partial V}
\right)_{N,\nu}
=
 B^2 \left(
\frac{\partial^2\varepsilon}{\partial B^2}
\right)_\nu,
\end{equation}
where $V$ means area. Here the ``internal pressure'' $P_{int} = P - BM/V$  excludes the pressure contribution from the magnetization $M$ due to edge current\cite{emresp2, internalpressure}. Note that the derivative is taken at fixed $\nu$ instead of fixed $B$. Thus, the internal compressibility has a finite value.  

In previous literatures\cite{emresp1,emresp2}, the explicit frequency dependence in (\ref{cv}) was not given, as for the discussion there it was not important. But from the relation between susceptibility and conductivity\cite{wenbook}, obtained from Kubo formulae (\ref{susceptibility}) (\ref{conductivity}) and particle conservation (\ref{ncons}),
\begin{equation}\label{sc}
\chi(\bbq,\omega) = \frac{q_\alpha q_\beta}{i e^2\omega^+}\sigma_{\alpha\beta}(\bbq,\omega),
\end{equation}
we see that even if one only considers the susceptibility at zero-frequency, it reflects the linear-in-frequency part of the conductivity. As we shall see later, it is essential to incorporate the $\omega$-dependence in (\ref{cv}) before taking the limit.
Explicitly, (\ref{cv}) and (\ref{sc}) shows the susceptibility-viscosity relation 
\begin{eqnarray}
\nonumber
\chi^{(4)}(\bbq,\omega) &=& \frac{q^4}{m^2(\omega^2-\omega_c^2)^2}
\left[
 \kappa_{int}^{-1} -
2\omega_c\eta_H(\omega) 
\right.
\\ \label{svw}
&& \qquad\quad\left. +\zeta(\omega) \frac{\omega}{i}+ \eta^S(\omega) \frac{\omega^2+\omega_c^2}{i\omega}
\right].
\end{eqnarray}
And at zero frequency $\omega=0$, the static susceptibility is
\begin{eqnarray}\nonumber
\chi^{(4)}(\bbq,\omega=0) &=& \frac{q^4}{m^2\omega_c^4}\left[
\kappa_{int}^{-1}-2\omega_c\eta_H(\omega=0)
\right.
\\ \label{sv}
&& \quad
+\omega_c^2 \lim_{\omega\rightarrow0} \left( \frac{\eta^S(\omega)}{i\omega}\right) \left.
-i \lim_{\omega\rightarrow0}\left(\omega\zeta(\omega)\right)
\right].\quad
\end{eqnarray}
 We stress that
these relations are obtained under the only assumption of Galilean invariance.  Thus,  the Hall viscosity is guaranteed to show up in the $q^4$ term in the density response function for a wide range of systems.

\section{Example: Integer Quantum Hall Systems}
We apply the above results to discuss the paradigm example of integer quantum Hall effect. The Hamiltonian involves only the kinetic part $\hat{H}_0 = \sum_i \hat\pi_\mu^i \hat\pi_\mu^i/2m = \hbar\omega_c\sum_i (\hat a_i^\dagger \hat a_i+1/2)$, where the ladder operator for Landau levels is $\hat{a}_i=(\hat\pi^i_x+ i\hat \pi^i_y)/ \sqrt{2\hbar eB}$, with the commutation relation $[\hat{a}_i,\hat{a}_j^\dagger]=\delta_{ij}$. Working in the spherical gauge $\hat{\mathbf{A}}^i  = \frac{B}{2}(-\hat{y}^i,\hat{x}^i,0)$, and define $\hat{\mathbf{R}}^i=\hat{\mathbf{p}}^i+e\hat{\mathbf{A}}^i$, we can introduce $\hat{b}_i=(\hat{R}_x^i -i\hat{R}_y^i)/\sqrt{2\hbar eB}$, with $[\hat{b}_i,\hat{b}_j^\dagger]=\delta_{ij}$. It specifies the degeneracy within each Landau level in terms of angular momentum $\hat{L}_z = \hbar\sum_i(\hat{b}_i^\dagger \hat{b}_i - \hat{a}_i^\dagger \hat{a}_i)$. Consider a ground state with the lowest $\nu$ Landau levels fully filled; each level has degeneracy $V/2\pi l_B^2$. 
Using equation (\ref{susceptibility}) we calculate the leading terms in susceptibility directly
\begin{eqnarray}\label{iqhechi2}
\chi^{(2)}(\bbq,\omega) &=& (ql_B)^2 \frac{\bar{n}\omega_c}{\hbar((\omega^+)^2 -\omega_c^2)},\\
\label{iqhechi4}
\chi^{(4)}(\bbq,\omega) &=& (ql_B)^4 \frac{\varepsilon}{\hbar^2}\left( \frac{1}{(\omega^+)^2-4\omega_c^2} - \frac{1}{(\omega^+)^2 - \omega_c^2}\right).
\end{eqnarray}
Here the energy density is $\varepsilon=(eB\nu)^2/4\pi m$, and the average density is $\bar{n}=\nu/2\pi l_B^2$, where the magnetic length $l_B^2 = \hbar/eB$.

  Next we use (\ref{sc}) (\ref{svw}) to calculate the same terms in susceptibility so as to verify the susceptibility-viscosity relations. 
The zeroth order term in conductivity is directly obtained from the Kubo formula (\ref{conductivity}), 
$
\sigma_{\mu\nu}^{(0)} (\omega)= \frac{ie^2\bar{n}}{m\omega^+} \frac{\omega^2}{(\omega^+)^2-\omega_c^2} \delta_{\mu\nu} + \frac{\bar{n} e^2\omega_c}{m(\omega_c^2-\omega^2)} \epsilon_{\mu\nu} ,
$
which reduces to the familiar one
$\sigma_{\mu\nu}^{(0)}(\omega=0) = e^2\nu/h $ at zero frequency. Using (\ref{sc}) we obtain (\ref{iqhechi2}).
Further\cite{emresp2},
$
\kappa_{int}^{-1} = 2\varepsilon,
\zeta(\omega) = 0,
\eta^S(\omega) = \frac{i\omega^+\varepsilon}{(\omega^+)^2 - 4\omega_c^2}, 
\eta_H(\omega) =  \frac{2\omega_c\varepsilon}{4\omega_c^2 - (\omega^+)^2 }.
$
 Then from (\ref{svw}) we have (\ref{iqhechi4}), as expected.

 
 We can further obtain the dynamic structure factor $S(\bbq,\omega)$ \cite{pn} using  the fluctuation-dissipation theorem Im$\chi(\bbq,\omega)= -\frac{\pi}{\hbar V} \left( S(\bbq,\omega) - S(-\bbq, -\omega) \right)$, and the identity $\frac{1}{x\pm i \epsilon} = {\cal P} \frac{1}{x} \mp \pi i \delta(x)$, 
 \begin{eqnarray}
 S^{(2)}(\bbq,\omega) &=& (ql_B)^2 \frac{N}{2} \delta(\omega-\omega_c),\\
 S^{(4)}(\bbq,\omega) &=& (ql_B)^4 \frac{N\nu}{8} [\delta(\omega-2\omega_c) - 2\delta(\omega-\omega_c)].
 \end{eqnarray}
 The static structure factor \cite{pn} $S(\bbq) = (1/N) \int_0^\infty d\omega \, S(\bbq,\omega)= \langle \hat{n}(\bbq)\hat{n}(-\bbq)\rangle $ is then
 \begin{equation}
 S(\bbq) = \frac{(ql_B)^2}{2} - \frac{\nu}{8} (ql_B)^4 + {\cal O}((ql_B)^6).
 \end{equation}
 For $\nu=1$, it reduces to the first two terms in the expansion of the well-known result  $S(\bbq) = 1- e^{- (ql_B)^2/2}$ for a system with fully-filled lowest Landau level\cite{gmp1986}.
 
\section{ Hydrodynamic analysis}
Having performed the strict microscopic analysis, it is interesting to see whether there is a macroscopic derivation  of the susceptibility-viscosity relations (\ref{svw}) (\ref{sv}), which would provide a simpler and more intuitive way to understand them. To this end, we next show that (\ref{svw}) (\ref{sv}) can be reproduces by pure classical hydrodynamic equations, despite the underlying highly quantum structure of the electronic liquid.  Such a method has been adopted to discuss the fractional quantum Hall liquids and Bose-Einstein condensates recently\cite{hydrorev,wiegmannhydro,becbook}.


In hydrodynamics, the microscopic details are averaged over, giving a few effective macroscopic variables,
\begin{equation}
n(\bbr,t) = \bar{n} + \delta n(\bbr,t), \qquad
\mathbf{u}(\bbr,t) =  \bar{\mathbf{u}} + \mathbf{v}(\bbr,t).
\end{equation}
Here $n, \mathbf{u}$ are the macroscopically averaged number density and the velocity respectively, including the equilibrium value $\bar{n}, \bar{\mathbf{u}}$, and the small deviation $\delta n(\bbr,t), \mathbf{v}(\bbr,t)$ caused by external perturbation. The dynamics is given by the Navier-Stokes equation (momentum conservation) (\ref{pcons}),
where the momentum $\mathbf{p}=mn\mathbf{u}$. The force comes from the Lorentz force of the uniform magnetic field and the external potential perturbation $\mathbf{f}=-enB\mathbf{e}_z\times \mathbf{u}-n\nabla\varphi(\bbr,t)$. The stress tensor is
 \begin{equation}
 T_{\mu\nu} = P_{int}\delta_{\mu\nu} + mnu_\mu u_\nu -\delta T_{\mu\nu},
 \end{equation}
where the viscous part $\delta T_{\mu\nu}$ is given by (\ref{stressdef})-(\ref{hallvisc}), and $P_{int}$ is the internal pressure introduced earlier. We stress that $P_{int}$ excludes the contribution to pressure from Lorentz force exerting on edge current, and therefore is suitable to be used here, as the Naivier-Stokes equation only concerns bulk properties and assumes no boundary effect. Consider the linear response regime where $\delta n(\bbr,t), \mathbf{v}(\bbr,t)$ are kept up to linear order, and use the equilibrium value $\bar{n}=\mbox{constant}, \bar{\mathbf{u}}=0$, the Navier-Stokes equation (\ref{pcons}) becomes
\begin{equation}\label{hydroeom}
m\bar{n}\partial_t \mathbf{v} = (e\bar{n}B+ \eta^H\Delta)( \mathbf{v}\times \mathbf{e}_z) + \nabla (\zeta\nabla\cdot\mathbf{v} - \bar{n}\varphi - P_{int} ) + \eta^S \Delta \mathbf{ v},
\end{equation}
On the other hand, the particle conservation, also kept up to linear order, gives
\begin{equation}
\partial_t\delta n + \bar{n}\nabla\cdot 
\mathbf{v} = 0,
\end{equation}
Then applying the alternative expression for compressibility $\kappa_{int} = \frac{1}{n}\frac{\partial n}{\partial P_{int}}$, which gives $\nabla P_{int} = \kappa_{int}^{-1}(\nabla \delta n)/\bar{n}$,
we have the density response to potential perturbation $\delta n = \chi \varphi$, where the susceptibility (in momentum space) reads
\begin{equation}\label{hydrochi}
\chi(\bbq,\omega) = \dfrac{\bar{n}^2q^2}{m \bar{n}\omega^2 - 
\left(\dfrac{\kappa_{int}^{-1}}{i\omega}- \eta^S - \zeta\right)i\omega q^2 + \dfrac{(m\bar{n}\omega_c-\eta^H q^2)^2}{\eta^Sq^2/i\omega - m\bar{n}}
}.
\end{equation}
In the low momentum expansion, the $q^2$ term is exactly given by  that for integer quantum Hall effect (\ref{iqhechi2}). This is because the intra-Landau level excitations start from the $q^4$ term\cite{gmp1986}. Thus, the $q^2$ contribution must entirely come from inter-Landau level excitations, whose characters are captured by the integer quantum Hall effect. The $q^4$ term is given by the same equation (\ref{svw}), as being derived from the Kubo formula. Thus, in the low momentum regime, the classical hydrodynamics  reproduces the strict susceptibility-viscosity relations obtained from Kubo formulae.

(In general, the viscosity coefficients $\zeta,\eta^S,\eta_H$ would have momentum dependence also\cite{dencurv1,dencurv2}. In the expansion of (\ref{hydrochi}) discussed above, we have taken the viscosity coefficients  $\zeta,\eta^S,\eta_H$ to be constant (only depends on frequency). This is suffficient when we expand $\chi(q,\omega)$ up to $q^4$ terms, as can be seen from (\ref{hydrochi}).  But when applying (\ref{hydrochi}) to higher orders in $q$, one needs to first expand $\zeta,\eta^S,\eta_H$ to higher powers in $q$  (i.e. $q^0$ and $q^2$ terms for $\zeta, \eta^S,\eta_H$) before expanding  (\ref{hydrochi}) (i.e. to $q^6$ term for $\chi(q,\omega)$)).

\section{Measuring Hall Viscosity}
The  relations (\ref{svw}), (\ref{sv}) connects the Hall viscosity with the susceptibility, which can be measured by scattering experiments. Since many recent works only concern the zero frequency value of Hall viscosity $\eta_H(\omega=0)$, we focus on equation (\ref{sv}) here. Note that in most cases the bulk viscosity $\zeta(\omega)$ does not diverge\cite{emresp2} at zero frequency. Then the last term in equation (\ref{sv}) vanishes and the formula can be rewritten as
  \begin{equation}   \label{measurehall}
\eta_H= 
  \frac{\kappa_{int}^{-1}}{2\omega_c}
  + \frac{\omega_c}{2} \lim_{\omega\rightarrow0} \left(\frac{\eta^S}{i\omega}\right)
   -\frac{m^2\omega_c^3}{2} \frac{\chi^{(4)}(q)}{q^4}.
  \end{equation} 
  In order to extract $\eta_H$ from the measurement of $\chi^{(4)}(q)$, one has to determine the first two terms in (\ref{measurehall}). The inverse internal compressibility $\kappa_{int}^{-1}=B^2(\partial^2\varepsilon(B)/\partial B^2)_\nu$ can be determined by the auxiliary measurement of magnetic susceptibility $\chi_M = \mu_0\left(\frac{\partial M}{\partial B}\right)_\nu = -\mu_0 \left(\frac{\partial^2\varepsilon}{\partial B^2}\right)_\nu$\cite{magneticchi} at constant filling fraction, or a local current measurement in response to the inhomogeneous magnetic field\cite{emresp1} $\delta\mathbf{j} = \nabla\times \delta\mathbf{M}  = -\varepsilon''(B)\mathbf{e}_z\times\nabla\delta B$. In particular, in high magnetic field where Landau level mixing is negligible, the interaction energy can be neglected compared with the kinetic energy. Then we have the free particle results $\varepsilon= \frac{(eB\nu)^2}{4\pi m}$ and $\varepsilon = \frac{(eB)^2\nu}{4\pi m}$ for integer and fractional quantum Hall systems respectively, giving $\kappa_{int}^{-1}=2\varepsilon$.


Further,  $\lim_{\omega\rightarrow0} (\eta^S/i\omega)$ generally yields a finite value, (See the  integer quantum Hall system for example), and needs careful evaluations.  Here we invoke the Kubo formula for shear viscosity\cite{emresp2}, 
$\eta^S = \frac{\hbar\omega^+}{V}\int_0^\infty e^{i\omega^+t } \langle [\hat{J}_{12}(t),\hat{J}_{12}(0)]\rangle_0$, to compute this quantity.  $\hat{J}_{12}$ is the off-diagonal element of the strain generator $\hat{J}_{\mu\nu}$. In the spectral representation,
\begin{equation}
\frac{\eta^S}{i\omega}
= 
\frac{2\hbar}{V} \sum_n \frac{\omega_{n0}|\langle 0| \hat{J}_{12}(0)|n\rangle|^2}{(\omega^+)^2-\omega_{n0}^2},
\end{equation}
where $\omega_{n0}=\omega_n - \omega_0$, with $\omega_n$ the frequency of energy  eigenstate $|n\rangle$. (In the presence of Landau level mixing, $|n\rangle$ also includes higher Landau level eigenstates). Therefore, if $\omega_{n0}=0$ or the matrix element vanishes, the corresponding term is zero; otherwise, we have the finite result
\begin{equation}\label{etas}
\lim_{\omega\rightarrow0} \left(
\frac{\eta^S}{i\omega}
\right) = -\frac{2\hbar}{V} {\sum_n}' \,\frac{|\langle 0| \hat{J}_{12}(0)|n\rangle|^2}{\omega_{n0}},
\end{equation}
where $\sum'$ means summing over energy eigenstates which are not degenerate with the ground state.
Explicitly, the off-diagonal element of the strain generator assumes the form
$\hat{J}_{12}(0) = \sum_i \left( -\{ \hat{x}^i, \hat{\pi}_y^i\}-eB\hat{x}^i\,^2 \right)/2\hbar$, and  can be written conveniently in the spherical gauge $\mathbf{\hat A}^i = \frac{1}{2}\mathbf{B}\times\hat{\bbr}^i$ as 
\begin{equation}\label{j12}
\hat{J}_{12}(0) = - \sum_i \frac{\hat x^i \hat p_y^i}{\hbar} = 
-\frac{1}{4} \sum_i [(\hat{a}_i - \hat{a}_i^\dagger)^2 
- (\hat{b}_i - \hat{b}_i^\dagger)^2].
\end{equation} 
where $\hat{a}_i, \hat{b}_i$ are inter- and intra-Landau level ladder operators defined previously. Consider, for instance, integer quantum Hall effect. Then the only non-zero matrix element in (\ref{etas}) is $\langle 0| \hat{a}_i^2 |2\rangle$, where $|2\rangle$ means the state with one electron from either of the top two filled Landau levels in the ground state being excited 2 levels upwards, corresponding to an excitation energy $\omega_{20} = 2\omega_c$. Thus, $\lim_{\omega\rightarrow 0} (\eta^S/i \omega) =  -\varepsilon/4\omega_c^2$, in consistency with the previous result. The evaluation for various fractional quantum Hall systems is left for future work, which generally requires a numerical evaluation of  (\ref{etas}) (\ref{j12}) in the presence or absence of Landau level mixings depending on specific experimental situations.
  
Thus, with $\kappa_{int}^{-1}$ and $\lim_{\omega\rightarrow0}(\eta^S/i\omega)$ being measured or calculated, one can use equation (\ref{measurehall}) to determine the Hall viscosity $\eta_H$ through the long wavelength part of static susceptibility $\chi^{(4)}(\bbq,\omega=0)$, which is related to scattering experiments\cite{pn}. Explicitly,  the quantity being measured is the dynamic structure factor $S(\bbq,\omega) = \sum_n S_n(\bbq) \delta(\omega-\omega_n(\bbq))$, where $\omega_n(\bbq)$'s are excitation modes. Using Kramers-Kronig relation Re$\chi(\bbq,\omega) = (1/\pi)\int_{-\infty}^\infty dz\, {\cal P} (\mbox{Im} \chi(\bbq,z))/(z-\omega)$ and fluctuation dissipation theorem, we have the static susceptibility $\chi(\bbq) = \chi(\bbq,\omega=0)$ as\cite{explainchisma}
\begin{equation}\label{chimeasure}
\chi(\bbq) = -\frac{2}{\hbar V} \sum_n \frac{S_n(\bbq)}{\omega_n(\bbq)}.
\end{equation}

The qualitative behavior of $\chi(\bbq)$ in small-q limit can be observed as follows. 
Using compressibility sum rule $\chi(\bbq\rightarrow0) = -N\kappa_T^{-1}$ and the incompressible feature of quantum Hall states, we see the constant term in $\chi(\bbq)$ always vanishes. Here $\kappa_T^{-1} = -V (\partial P/\partial V)_{N, B}$ is the ususal isothermal compressibility. Moreover, the low-lying excitations within the lowest Landau level correspond to $\chi(\bbq)$ starting from $q^4$ term\cite{explainchisma}. That means if the inter-Landau level excitation is supressed in the large magnetic field limit, (or we can cancel such contribution by subtracting the $\chi^{(2)}$ given by equation (\ref{iqhechi2})),  $\chi^{(4)}$ is the leading order in $\chi(\bbq)$.  In sum, the algebraic relation (\ref{chimeasure}) clearly means that it will be sufficient to extract the $q^4$ term in $\chi(\bbq)$ by measuring only the long wavelength part of $S(\bbq,\omega)$, as is usually the case in current experiments.

\section{Conclusion and outlook}
The relation between susceptibility and viscosity (\ref{svw}) (\ref{sv}) is presented and discussed, showing the role of viscosity coefficients in the $q^4$ term of susceptibility  for a general class of Galilean invariant system. It suggests the possibility of measuring the Hall viscosity in terms of density response experiments for a wide range of quantum Hall systems. 

In addition, it is worth mentioning the connection of this work with cold atom systems, where  quantum Hall states can be simulated using rotating atomic gases\cite{markus}. The quantum Hall state is approached when the rotation frequency of the trapped gas approaches the critical frequency, where the centrifugal force almost cancels the trapping force, and the Galilean invariance is approximately satisfied. Since in cold atom systems, the density measurement and the potential engineering are the most standard experimental tools, and can be made to high accuracy, our analysis paves the way for further discussions of Hall viscosity signatures in cold atom experiments.


\section{ Acknowledgement}
I  thank Tin-Lun Ho, Yuan-Ming Lu,  and  Alexander Abanov  for  discussions, and especially T.-L. Ho for helpful comments and suggestions about the manuscript. This work is supported by NSF Grant DMR-1309615, and the NASA Grant 1501430.

\begin{thebibliography}{99}
\bibitem{magneticchi}
The usual convention is $\chi_M=\partial M/\partial H$; $B=\mu_0(1+\chi_M)H$. But except for ferromagnetic systems, $\chi_M\ll1$(QH systems are mostly paramagnetic). So we have $\chi_M=\mu_0 \partial M/\partial B$.
\end{thebibliography}
\end{document}